\documentclass[11pt,dvips]{article}

\usepackage{epsfig,times} 
\usepackage{picinpar}
\usepackage{wrapfig}
\usepackage{floatflt}
\makeatletter
\renewcommand{\section}{\@startsection{section}{1}{0in}
        {0.4\baselineskip}{0.1\baselineskip}{\Large\bf}}
\renewcommand{\subsection}{\@startsection{subsection}{2}{0in}
        {0.25\baselineskip}{-\baselineskip}{\large\bf}}
\renewcommand{\subsubsection}{\@startsection{subsubsection}{3}{0in}
        {0.1\baselineskip}{-\baselineskip}{\normalsize\bf}}
\makeatother
\begin{document}

%
%  Session and Paper Code:
%\thispagestyle{myheadings}
%
%  ***INSTRUCTIONS:***  Replace `OG 9.9.9' in the command argument below
%                       with your assigned session and paper code:
%\markright{HE 5.1.03}
%
\makeatletter\newcommand{\ps@icrc}{
\renewcommand{\@oddhead}{\slshape{HE.5.1.03}\hfil}}
\makeatother\thispagestyle{icrc}
%
%  ***INSTRUCTIONS:***  Replace `OG 9.9.9' in the command argument below
%                       with your assigned session and paper code:
%\markright{OG 9.9.9}
%

%  Title:
\begin{center}
%
%  ***INSTRUCTIONS:***  Replace `Instructions for Preparation of Manuscript'
%                       with your paper's title:
{\LARGE \bf Search for WIMPS using Upward-Going Muons in MACRO}
\end{center}

%  Author List:
\begin{center}
%
%  ***INSTRUCTIONS:***  Replace authors and addresses below with your own:
%
{\bf T. Montaruli$^{1}$ for the MACRO Collaboration}\\
{\it $^{1}$Istituto Nazionale di Fisica Nucleare, Bari, I-70126, Italy\\}
\end{center}

%  Abstract:
\begin{center}
{\large \bf Abstract\\}
\end{center}
\vspace{-0.5ex}
%
%  ***INSTRUCTIONS:***  Replace text below with your own abstract:
%
We present updated results on the search for a neutrino signal from the 
core of the Earth and of the Sun induced by Weakly Interacting Massive
Particles (WIMPs). In this paper we concentrate on
neutralinos as WIMP candidates.
The 971 and 642 events used respectively for the search from the Sun 
and from the Earth are compatible with the background of atmospheric 
neutrinos. 
Consequently we calculate flux limits for various search cones 
around these sources.
Limits as a function of the neutralino mass are given and compared to
the supersymmetric (SUSY) models.
%interpretation of the DAMA/NaI annual modulation indication.
%
%  Leave this line skip in place:
\vspace{1ex}

%
%  Manuscript text:
%
\section{WIMPs from the Earth and the Sun:}
\label{sec:intro}
We describe an indirect method to look for non-baryonic dark 
matter WIMPs.
The best WIMP candidate is the lightest SUSY particle
(LSP) which in the Minimal Supersymmetric Standard Model (MSSM) is
expected to be stable if R-parity is conserved
and hence should be present in the Universe as a cosmological relic 
from the Big Bang. The linear combination of Higgsinos and gauginos,
the neutralino $\chi$, is currently considered 
the best candidate for cold dark matter since its couplings and 
mass range naturally give the relic density required to
explain halo dark matter.
Neutralinos are described by 3 parameters in the MSSM assuming
the GUT relation between gaugino masses: one of the gaugino masses, 
%$M_{1}$ or $M_{2}$, 
the Higgsino mass parameter $\mu$ and 
the ratio of the Higgs doublet 
vacuum expectation values $\tan\beta$. Moreover, if universality 
is assumed, the MSSM phenomenology is
described by these parameters, 
the universal trilinear scalar coupling $A$, 
the degenerate scalar mass $m_{0}$ and 
the mass of the pseudoscalar neutral Higgs $m_{A}$.
Experimental searches at LEP
%, without further theoretical assumptions,
set a lower limit on $m_{\chi}$ at $\sim 32$ GeV 
and suggest an upper limit at $\sim 600$ GeV 
if one requires that the neutralino cosmological abundance
$\Omega_{\chi} h^{2} \le 0.3$ (Ellis, 1999). This upper limit is not 
yet achievable by the forthcoming LEP and Tevatron runs.
Direct and indirect searches at underground detectors
explore SUSY parameters
and are complementary to the future LHC measurements.
Direct searches look for a signature of a direct 
scattering of a WIMP from a nucleus in the detector. The DAMA
experiment ($\sim 100$ Kg NaI(Tl))
sees an indication of an annual modulation of the rate
which could be due to the Earth's motion around the Sun and 
the change of the Earth's velocity relative to the incident WIMP. 
%Bernabei et al., 1999, have reported an annual modulation 
%signal with high statistical significance in the counting rates they 
%observe in well-shielded, highly radiopure NaI detectors.
The 19511 kg day data favor at 99.6$\%$ c.l. the presence of
an annual modulation signal which, if interpreted in terms of
WIMPs, implies a mass of $m_{W} = (59^{+17}_{-14}$) 
GeV (Bernabei et al., 1999). 
This indication should be checked using different
techniques, such as the indirect detection of trapped WIMPs inside the
core of the Earth and of the Sun. The signature would be an excess of neutrino
events resulting from WIMP-WIMP annihilations 
around the direction of the vertical of the apparatus and of the 
Sun beyond the known atmospheric $\nu$ background (Jungman, Kamionkowski \&
Griest, 1996).
MACRO measures neutrinos indirectly as upward-going muons and
has presented results of the WIMP search in Ambrosio et al., 
1998a, to which we refer for details. 
We update this search including the data
collected during Mar. 98-Feb. 99.
\section{MACRO Updated Results on WIMPs:}
The MACRO detector at the Gran Sasso Laboratories, with
overall dimensions of $12 \times 76.6 \times 9$ m$^3$, detects
upward-going muons through the time-of-flight measurement using 600 tons of
liquid scintillator inside 12 m long boxes (time resolution $\sim 500$ psec). 
A system of around $20,000$ m$^2$
of streamer tubes reconstructs tracks with angular resolution $\le 1^{\circ}$.
The lower part of the apparatus is filled
with rock absorber setting a 1 GeV threshold for vertical $\mu$s.
The upward-going muon measurement relative to the construction period 
of MACRO (Mar. 89 - Apr. 94: 1.38 yr of running of 1/6 of the lower
apparatus and 0.41 yr of the lower detector, inefficiencies included) 
is described in Ahlen et al., 1995. 
Since then, MACRO is in its full configuration 
(3.93 yr, inefficiencies included) and further results are
in Ambrosio et al., 1998a and Ambrosio et al., 1998b.

For the WIMP search for the Earth we use the sample of 642 throughgoing upward
muons selected with the requirement that the track crosses at least 200 
g/cm$^2$ in the MACRO rock absorber, 
which reduces the background due to soft
$\pi$s produced at large angles by downward-going $\mu$s to $\sim 1\%$
(Ambrosio et al., 1998c).
Releasing this cut, we use 971 upward-going $\mu$s for the search for the Sun 
because background rejection is not so critical for moving sources and the
increase in exposure offsets the slight increase in background.
\begin{tabwindow}[1,r,%
{\mbox{
\begin{tabular}{|cccccccc|} \hline
 \multicolumn{1}{|c}{ }&
 \multicolumn{4}{c}{EARTH}&
 \multicolumn{3}{c|}{SUN} \\ \hline
 \multicolumn{1}{|c}{Half-}  
&\multicolumn{1}{c}{Data}
&\multicolumn{1}{c}{Back-} 
&\multicolumn{1}{c}{Norm.}
&\multicolumn{1}{c}{Flux Limit} 
&\multicolumn{1}{c}{Data} 
&\multicolumn{1}{c}{Back-} 
&\multicolumn{1}{c|}{Flux Limit} \\ 
cone & & ground& factor & ($E_{\mu} > 1.5$ GeV) & 
& ground& ($E_{\mu} > 2$ GeV)\\
 & & events&        & (cm$^{-2}$ s$^{-1}$)  & & events& (cm$^{-2}$ s$^{-1}$)\\
\hline
$30^{\circ}$ & 102 & 150.2 & 0.83 & 2.01 $\times 10^{-14}$& 69 &
58.9 & 6.38 $\times 10^{-14}$\\
$24^{\circ}$ & 70  &  96.2 & 0.80 & 1.56 $\times 10^{-14}$& 41 & 
37.3 & 4.06 $\times 10^{-14}$ \\
$18^{\circ}$ & 44  &  53.0 & 0.78 & 1.28 $\times 10^{-14}$& 22 & 
20.9 & 2.77 $\times 10^{-14}$\\
$15^{\circ}$ & 32  &  36.8 & 0.77 & 1.03 $\times 10^{-14}$& 14 & 
14.5 & 2.07 $\times 10^{-14}$\\
$9^{\circ} $ & 12  &  13.7 & 0.77 & 6.58 $\times 10^{-15}$ & 5 &  
5.3  & 1.42 $\times 10^{-14}$\\
$6^{\circ}$  & 4   &   6.2 & 0.77 & 5.07 $\times 10^{-15}$& 2 &   
2.3  & 1.07 $\times 10^{-14}$\\
$3^{\circ}$  & 0   &   1.6 & 0.77 & 2.89 $\times 10^{-15}$& 2 &  
0.6  & 1.35 $\times 10^{-14}$\\ \hline
\end{tabular}
\label{tab1} 
}},%
{Observed and atmospheric $\nu$-induced background and 90$\%$
c.l. $\mu$ flux limits as a function of half-cone angles around the Earth core
and the Sun.
For the Earth, the expected background events 
are multiplied by the ratio of observed to expected events outside each cone. 
The Earth results are for the no oscillation scenario. 
The average exposure for the Earth is 3272 m$^{2}$ yr and for the Sun
1116 m$^{2}$ yr.}]
For the Earth, the expected background due to interactions of 
atmospheric $\nu$s in the rock below MACRO is evaluated
with a full Monte Carlo described in Ambrosio et al., 1998b
using the Bartol $\nu$ flux (Agrawal et al., 1996), the GRV(94) 
DIS parton distributions (Gl\"uck,\\
Reya \& Vogt, 1995) 
and the muon energy loss
as in Lohmann et al., 1985. We estimate a total uncertainty in the 
calculation of upward-going muon fluxes of 17$\%$. For the Earth we have
even considered a $\nu_{\mu} \rightarrow \nu_{\tau}$ 
oscillation scenario with parameters $\Delta m^{2} = 0.0025$ eV$^2$,
$\sin^{2}2\Theta = 1$ as suggested by the flux measurement
reported at this conference (Ronga et al., 1999).
For the Earth search, the expected number of atmospheric induced events
in the no oscillation scenario is (including the contribution due to 
$\nu$-interactions inside the bottom part of the apparatus which are
selected as throughgoing muons) $835 \pm 142$ and in the oscillation 
scenario $581 \pm 99$.

$~~~~~$For the Sun we have compared the 971 measured events with a 
different simulation with respect to that used for the Earth. 
This is obtained by mixing randomly the local coordinates of measured
upward-going events and times gathered during the entire data-taking.
This method takes into account the contribution of events
produced by internal $\nu$-interactions in the MACRO absorber.
In Fig. 1(a) we show the angular distributions 
of the measured and expected
events from atmospheric neutrinos for the Earth and the Sun. For the Sun this
distribution depends strongly on time due to the
motion of the Sun. In Fig. 1(b) we show the muon
flux limits for 10 search cones from 3$^{\circ}$ to 10$^{\circ}$.
In the case of the Earth, the expectation from atmospheric
$\nu$s in the region of interest for the signal is larger than the data; 
we then evaluate flux limits multiplying the
expected number of events by the ratio of the data to the expectation
outside the cone where we look for the signal. This normalization 
is motivated by the high uncertainty in the normalization of upward-going
muon flux calculations, whereas the shape error in the flux distribution is
a few percent only.  
Moreover, since the number of detected events is less than the normalized 
expected events, we set conservative flux limits assuming that 
the number of measured events equals the number of expected ones 
(Caso et al., 1998). 
With this normalization, 
Earth limits considering $\nu$-oscillations agree with 
the ones in the case of no oscillations within 7$\%$.
On the other hand, for the Sun, 
having used data to evaluate the expected numbers from
atmospheric $\nu$s, oscillation effects are automatically included
in the given limits.
In Table 1 we show measured and expected events and flux
limits for some of the cones 
calculated assuming a minimum $\mu$ energy of 1.5 GeV and 2 GeV for the
Earth and the Sun, respectively. The minimum energy for the Sun is higher
because tracks pointing toward it are more slanted than vertical
tracks pointing to the core of the Earth and hence cross a larger amount of
MACRO absorber.
The average exposures (live-time times detector
area in the direction of the expected signal from the source of WIMP
annihilation) for cones between $3^{\circ}$ 
and $30^{\circ}$ is 3272 m$^{2}$ yr for the Earth
and 1116 m$^{2}$ yr for the Sun. 
We estimate 
a maximum error of 5$\%$ on flux limits assuming these 
minimum energies for flux limit calculation with respect to a calculation
which takes into account the dependence of the acceptance of the
apparatus and of the neutrino fluxes from $\chi-\bar{\chi}$ annihilation. 
\end{tabwindow}
%\begin{figure}                   
\begin{tabular}{cc}
\epsfig{file=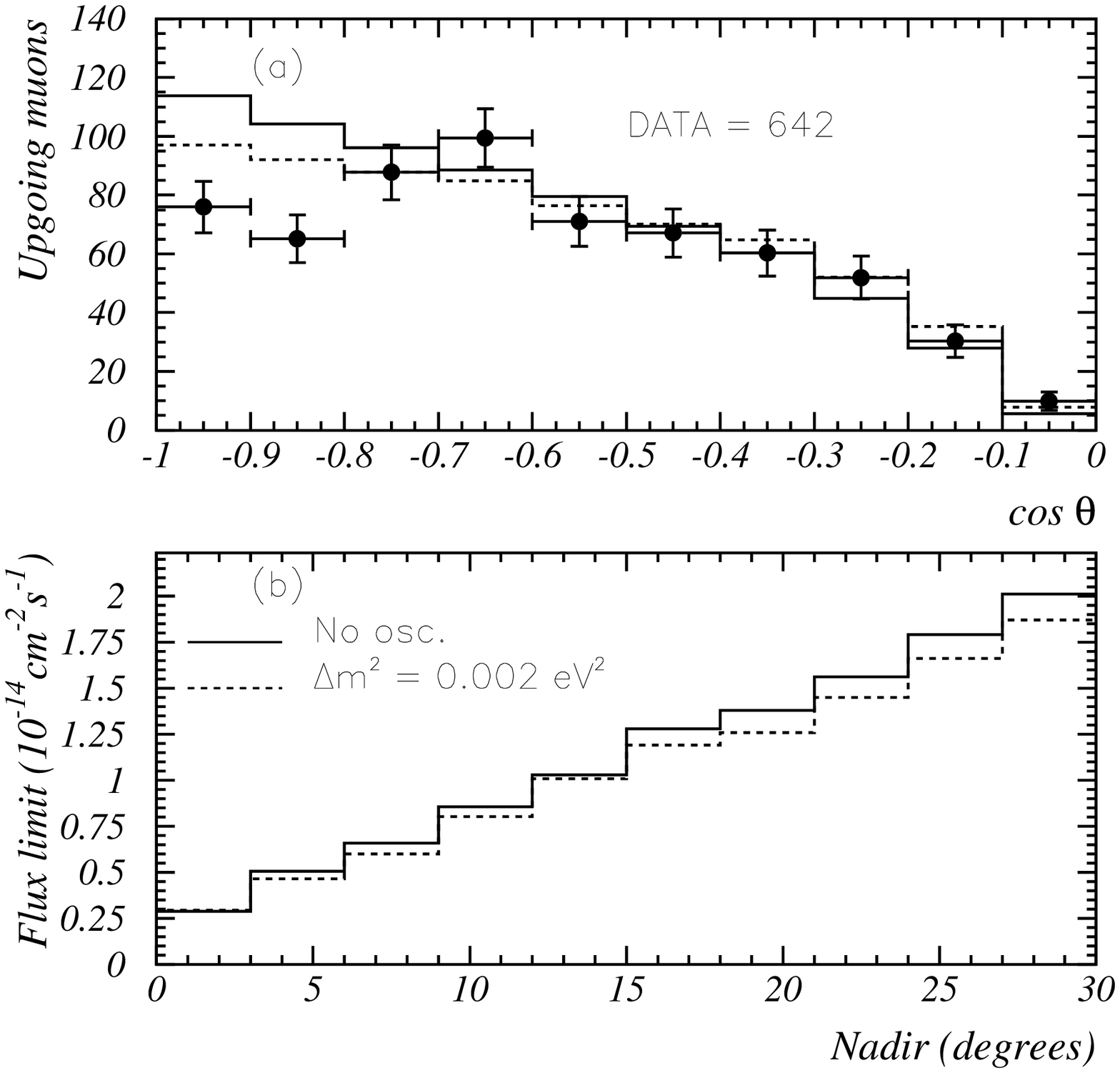,width=8.cm,height=9.5cm}&
\epsfig{file=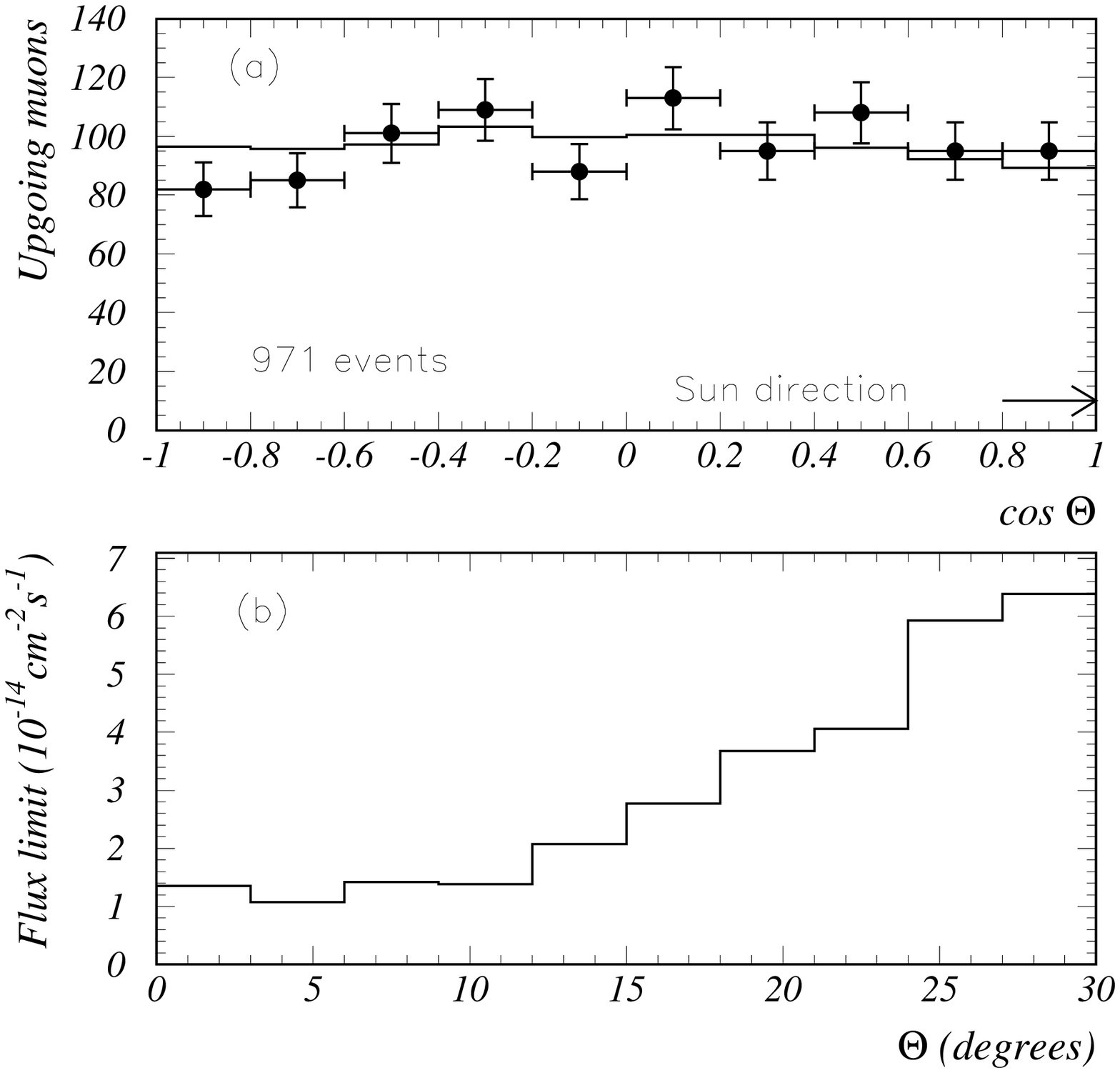,width=8.cm,height=9.5cm}
\end{tabular}
%\caption{
\textbf{Figure 1: }(a) On the left: zenith angular distribution 
of measured and expected events
for the Earth search. 
Solid and Dotted lines: expectations for atmospheric $\nu$s
for no-oscillations and $\nu_{\mu} \rightarrow \nu_{\tau}$
oscillations with $\Delta m^{2} = 0.0025$ eV$^2$; black circles: data.
The expected distributions are multiplied by the ratio $R$ of the measured
events over the expected ones outside the largest window $30^{\circ}$
($R = 0.83$ for the solid line
and 1.15 for the dashed line).
On the right:  distribution of the cosine of the angle from the Sun direction.
(b) $\mu$ flux limits (90$\%$ c.l.) vs. the
half-width of the cone from the source direction.
%}
%\label{fig1}
%\end{figure}
\vskip 0.5 cm
We calculate flux limits assuming the neutralino is a WIMP candidate.
We estimate search cones considering the angular distribution of upward-going
muons around the direction of neutrinos from neutralino annihilation which
mainly depends on the neutralino mass using Bottino et al., 1995 
flux calculations. We choose to calculate flux limits in those cones which 
include 90$\%$ of the expected signal. We estimate that extreme assumptions
on the annihilation channels (the final states are fermion pairs, 
gauge/Higgs bosons) 
produce a maximum variation of flux limits of about 17$\%$.
The Bottino models are shown as dots (model parameters are described in 
detail in Ambrosio et al., 1998b)
and circles in Fig. 2
as a function of neutralino mass for the Earth and the Sun and a 
minimum muon energy of 1 GeV. The solid lines represent MACRO flux limits
corrected for the 90$\%$ collection efficiency of the cones
and including correction factors to translate limits in Table 1 to the 1 GeV
threshold used in the calculation. These factors are maximum for low
neutralino masses (1$\%$ for the Earth and 10$\%$ for the Sun at $m_{\chi} =
60$ GeV). 
As shown in detail in Ambrosio et al., 1998b,
MACRO's experimental limit from the Earth rules out a considerable number
of SUSY configurations based on the interpretation of the DAMA/NaI
data (Bottino et al., 1999), even assuming neutrino oscillations of
the atmospheric $\nu$ background. It should anyway be considered that
even if the MACRO flux limits vary within 7$\%$ if the oscillation or 
no oscillation hypothesis is assumed, $\nu_{\mu}-\nu_{\tau}$
oscillations with the already quoted parameters could lower 
neutrino flux calculations from neutralino annihilation by at most 
a factor of two (Fornengo, 1999). 
In the case of the Sun, MACRO has less overlap in sensitivity for low
neutralino masses with direct searches because the indirect search is
more sensitive to spin-dependent scattering.   
%\begin{figure}
\begin{center}                   
\begin{tabular}{cc}
\epsfig{file=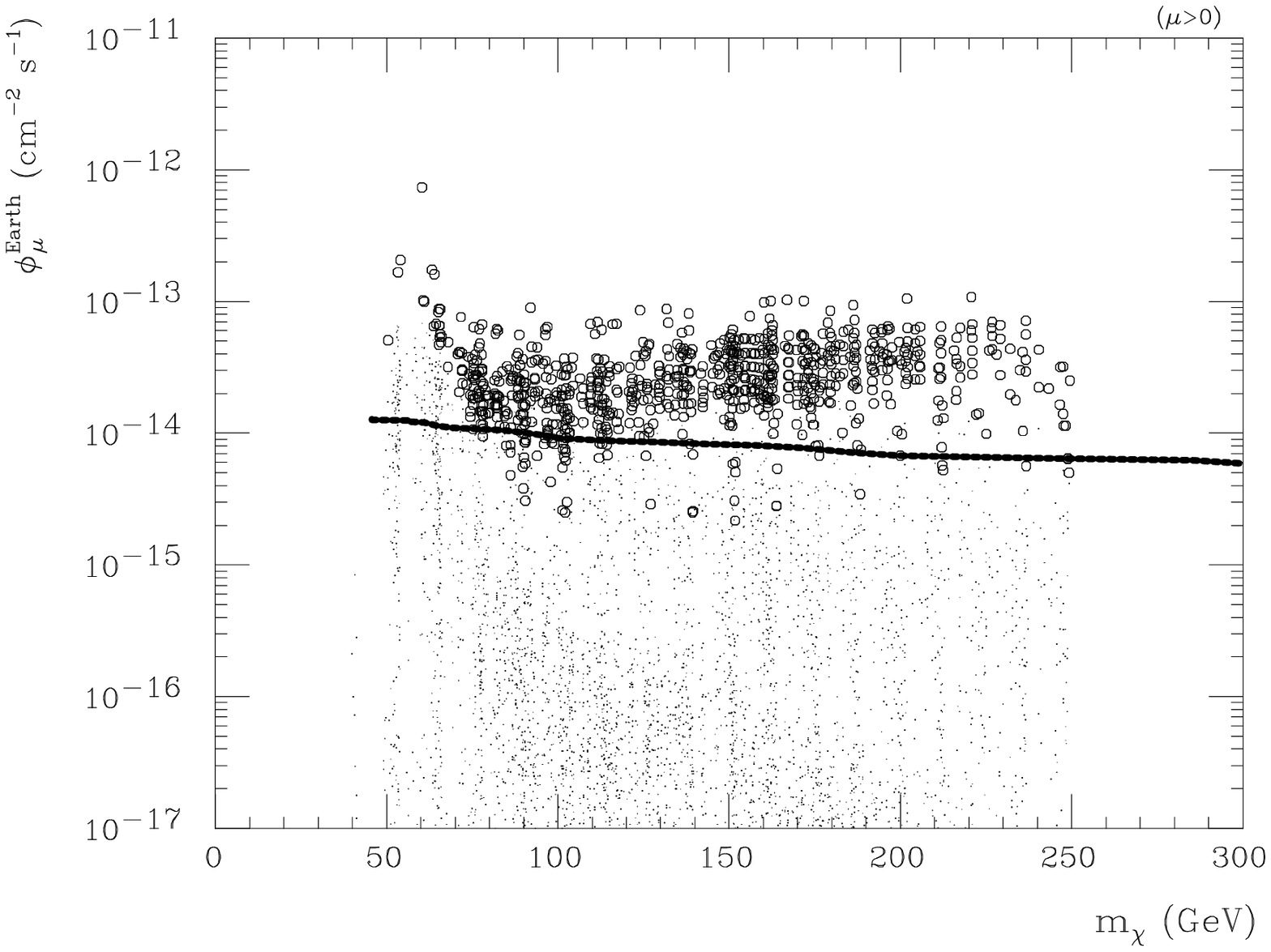,width=7.cm,height=8.5cm}&
\epsfig{file=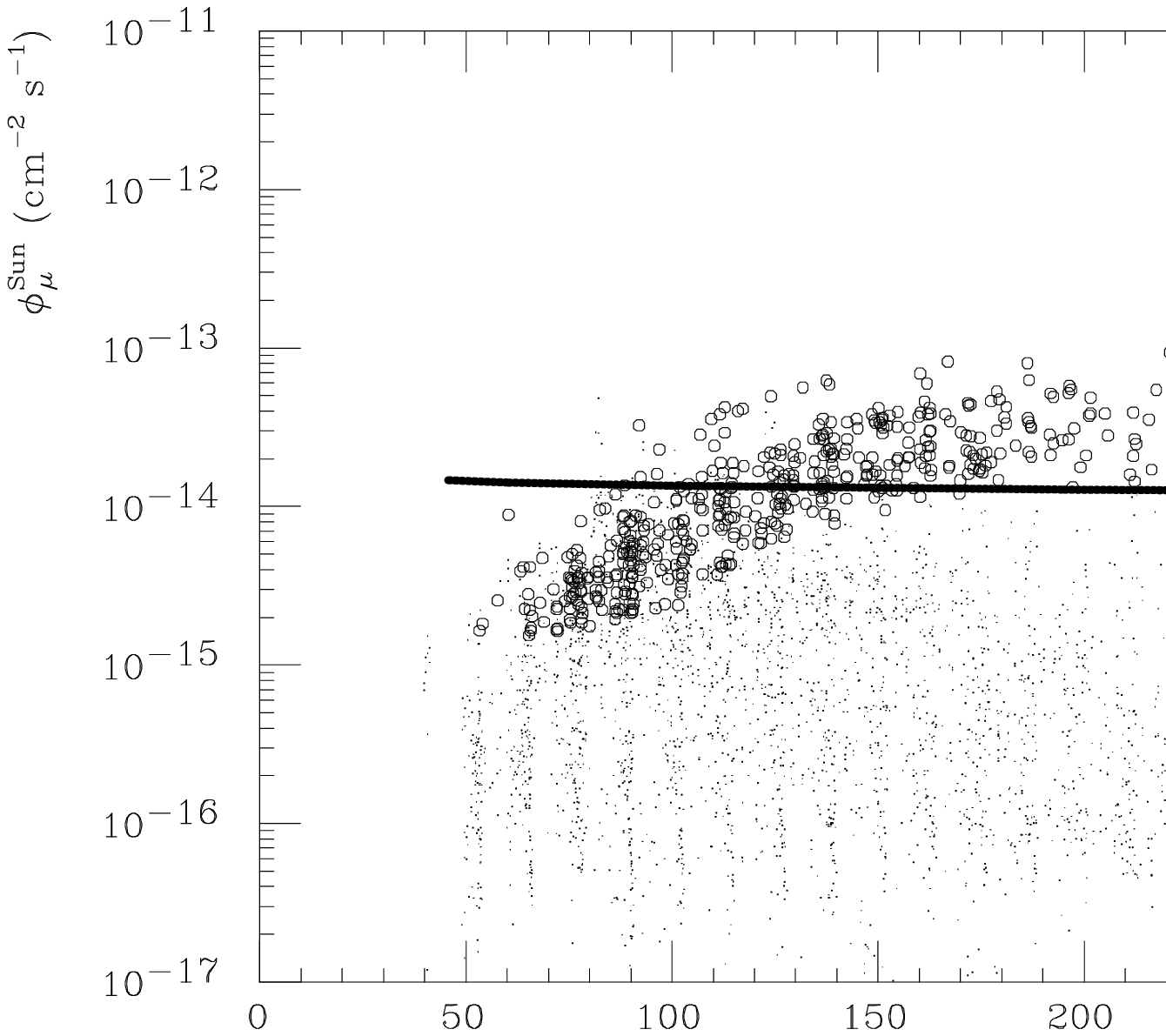,width=5.5cm,height=8.5cm}
\end{tabular}
%\caption{
\end{center}
\textbf{Figure 2: } On the left: dots are upward-going $\mu$ fluxes 
vs $m_{\chi}$ 
for $E_{\mu}^{th} = 1$ GeV from the Earth for various model parameters
(Bottino et al., 1995); solid line: MACRO flux limit (90$\%$ c.l.). 
Open circles: models excluded by direct measurements (Bernabei
et al., 1996).
On the right: the same as on the left for the Sun.
%}
%\label{fig2}
%\end{figure}
\vspace{1ex}
\begin{center}
{\Large\bf References}
\end{center}
%  ***INSTRUCTIONS:***  Enter your references alphabetically following the 
% format
%                       of the example citations below.
Agrawal, V., Gaisser, T.K., Lipari, P. \& Stanev, T., 1996, 
Phys. Rev. 53 1314\\
Ahlen, S., et al., MACRO Collaboration, 1995, Phys. Lett. B357, 481\\
Ambrosio, M., et al., MACRO Collaboration, 1998a, subm. to
Phys. Rev. D15 and hep-ex/9812020 \\
Ambrosio, M., et al., MACRO Collaboration, 1998b, Phys. Lett. B434, 451\\
Ambrosio, M., et al., MACRO Collaboration, 1998c, Astrop. Phys. 9, 105\\
%Ambrosio, M., et al., MACRO Collaboration, 1999, Phys. Rev. D59, 012003\\
Bernabei, R., et al., 1996, Phys. Lett. B389, 757\\
Bernabei, R., et al., 1999, Phys. Lett. B450, 448\\
Bottino, A., et al., 1995, Astrop. Phys. 3, 65 \\
Bottino, A., et al., 1999, Astrop. Phys. 10, 203\\ 
Caso, C., et al., Review of Particle Physics, 1998, 
Eur. Phys. J., C3, 1\\
Ellis, J., 1999, astro-ph/9903003\\
Fornengo, N, 1999, hep-ph/9904351\\
Gl\"uck, M., Reya, E. \& Vogt, A., 1995, Z. Phys. C67, 433\\
Jungman, G., Kamionkowski, M., \& Griest, K., 1996, 
Phys. Rep. 267, 195\\
Lohmann, W., et al., 1985, CERN Yellow Rep., CERN-EP/85-03\\
Ronga, F., for the MACRO Collaboration, 1999, these Proceedings\\
\end{document}